\documentclass[5p,times]{elsarticle}

\usepackage{textcomp,gensymb}
\usepackage{multirow}

\usepackage{float}
\usepackage[latin1]{inputenc}   
\usepackage{amsmath} 
\usepackage{epstopdf}           %
\usepackage{flushend}           %
\newcommand{\un}[1]{\ensuremath{\,\mathrm{#1}}}

\usepackage{lmodern}
\usepackage[T1]{fontenc}
\usepackage{xcolor}
\usepackage{upgreek}

\usepackage{amssymb}

\usepackage{lineno}

\usepackage[figuresright]{rotating}
\usepackage{color,soul}

\usepackage{subfigure}

\begin{document}
\begin{frontmatter}

\title{High Resolution 3D Strain and Orientation Mapping within a Grain of a Directed Energy Deposition Laser Additively Manufactured Superalloy}

\author[First,Second,Third,Seventh]{Y. Chen}
\author[Fourth]{Y.T. Tang}
\author[Fifth]{D.M. Collins}
\author[Sixth]{S.J. Clark}
\author[Second]{W. Ludwig}
\author[Second]{R. Rodriguez-Lamas}
\author[Second]{C. Detlefs}
\author[Fourth]{R.C. Reed}
\author[Third]{P.D. Lee}
\author[Seventh]{P.J. Withers}
\author[Second]{C. Yildirim\corref{cor1}}
\ead{can.yildirim@esrf.fr}

\cortext[cor1]{Corresponding Author}

\address[First]{School of Engineering, RMIT University, Melbourne, 3000, Australia}
\address[Second]{European Synchrotron Radiation Facility, 71 Avenue des Martyrs, CS40220, 38043 Grenoble Cedex 9, France.}
\address[Third]{Mechanical Engineering, University College London, Torrington Place, London WC1E 7JE, UK}
\address[Fourth]{Department of Materials, University of Oxford, Oxfordshire, UK}
\address[Fifth]{School of Metallurgy and Materials, University of Birmingham, Edgbaston, Birmingham B15 2TT, UK}
\address[Sixth]{X-ray Science Division, Advanced Photon Source, Argonne National Laboratory; Lemont, IL 60439
}
\address[Seventh]{Henry Royce Institute, Department of Materials, University of Manchester, Oxford Road, Manchester , UK}

\date{\today}

\begin{abstract}

The industrialization of Laser Additive Manufacturing (LAM) is challenged by the undesirable microstructures and high residual stresses originating from the fast and complex solidification process. Non-destructive assessment of the mechanical performance controlling deformation patterning is therefore critical. Here, we use Dark Field X-ray Microscopy (DFXM) to non-destructively map the 3D intragranular orientation and strain variations throughout a surface breaking grain within a directed energy deposition nickel superalloy. DFXM results reveal a highly heterogenous 3D microstructure in terms of the local orientation and lattice strain. The grain comprises $\approx$ 5\,$\upmu$m-sized cells with alternating strain states, as high as 5 $\times 10^{-3}$, and orientation differences <0.5$\degree$. The DFXM results are compared to Electron Backscatter Diffraction measurements of the same grain from its cut-off surface. We discuss the microstructure developments during LAM, rationalising the development of the deformation patterning from the extreme thermal gradients during processing and the susceptibility for solute segregation.

\end{abstract}


\end{frontmatter}
\begin{sloppypar}
Laser Additive Manufacturing (LAM) is a powerful and versatile technique for the direct layer-by-layer fabrication of complex 3D components from metallic alloy powders/wires \cite{gu2012laser} having the potential to revolutionize manufacturing processes. LAM is driven by component requirements such as unconventional geometries, small production numbers, digital manufacturing, and high-value-added applications in aerospace, power generation, and biomedical industries. Directed Energy Deposition LAM (DED-LAM) is amongst the most promising methods in LAM for industrial applications, allowing for building large freeform components and also offering \textit{in-situ} repairability. 
However, technological challenges restrict the mechanical performance of final products fabricated via LAM. The material experiences rapid cooling at $\approx $10$^{4}- 10^{6}$\,K/s, ca. $10^{5}\times$ faster than conventional processes (ca.1-10\,K/s) \cite{hooper2018melt}. The high cooling rate associated with laser processing can result in microstructural defects, introducing significant levels of residual stresses and high dislocation densities in the as-fabricated state. Highly-localized melting and non-equilibrium solidification dynamics creates residual stresses  on the order of >600\,MPa in 316L stainless steel \cite{fergani2017, li2018residual}. These stresses have a detrimental effect on the mechanical performance of the products, causing defects including cracking, delamination, and loss of dimensional accuracy. Therefore,  understanding the stress state of these complex microstructures is of significant technological importance.

A commonly-used technique for the characterization of AM microstructures is Electron Backscatter Diffraction (EBSD), which can quantify information about the structure \cite{WANG201768}, crystal orientation \cite{GOKCEKAYA2020101624}, phase \cite{ALAM2020110138} or strain \cite{honnige2017improvement} in the material. The measured diffraction signal typically comes from the surface 10-50\,nm of the sample. Combining with focused ion beam (FIB) layer-by-layer sectioning, it is possible to obtain 3D information of the microstructure. However, destructive specimen preparation (i.e. sectioning and polishing) may alter the microstructure near the surface and precludes dynamic studies in 3D. These methods inevitably suffer from surface relaxation of residual stresses, making it impossible to interpret the bulk behaviour of the stress distribution. The influence of such surface effects is difficult to quantify, making microstructure assessments with high confidence difficult.

Among non-destructive methods, synchrotron X-ray imaging has proved a powerful technique, capturing features such as porosity and regions where the powder was not fused in mm-sized samples. The technique allows for dynamic studies thanks to high image acquisition rates \cite{zhao2017real,leung2018situ, chen2021synchrotron}. Using X-ray micro-computed tomography ($\upmu$CT), non-destructive volumetric characterization of additively-manufactured parts can be achieved \cite{thompson2016x, wolff2021situ}. While these methods provide important insights, they do not provide information on the strain or crystal orientations of the material. Over the last 20 years, diffraction-based synchrotron imaging techniques such as 3D X-ray Diffraction (3DXRD)/High Energy Diffraction Microscope (HEDM)\cite{prithivirajan2021direct} and Diffraction Contrast Tomography (DCT)\cite{ludwig2008x} have been used extensively to map grain structures in 3D. Although these techniques provide grain-level orientation and strain in a given volume, they provide limited information on the intragranular level due to their spatial resolution ($\approx 1-2\,\upmu$m). Moreover, the strain and orientation spreads that are caused by the high cooling rates make 3D reconstruction difficult (due to diffraction peak-overlap problem in indexing). Non-destructive characterization of the strain and orientation state of the intragranular cellular microstructure of rapidly-cooled AM parts thereby remains a challenge. 

Dark Field X-ray Microscopy (DFXM), an emerging synchrotron imaging technique, can address the above-mentioned challenges. DFXM is a powerful diffraction-based technique for imaging strains and orientations in mm-sized crystalline materials with $\approx$ 150\,nm spatial and 0.001$\degree$ angular resolutions, respectively \cite{Simons2015,yildirim2020probing}. Using an objective lens analogous to dark field transmission electron microscope, it offers bulk sensitivity in probing the intragranular details for a grain of interest (GOI).

In this paper, we aim to unveil as-fabricated intragranular microstructures in a DED-LAM formed superalloy in order to build connections with the physico-chemical aspects of the rapid solidification during manufacturing. Using DFXM, we characterize the 3D orientation variation and strain distributions of grain from its cut-off surface to the fully embedded interior. We compare DFXM results with EBSD measurements of the \textit{same} grain. 


\begin{figure}[h!]
    \centering
    \resizebox{0.8\columnwidth}{!}
    {\includegraphics{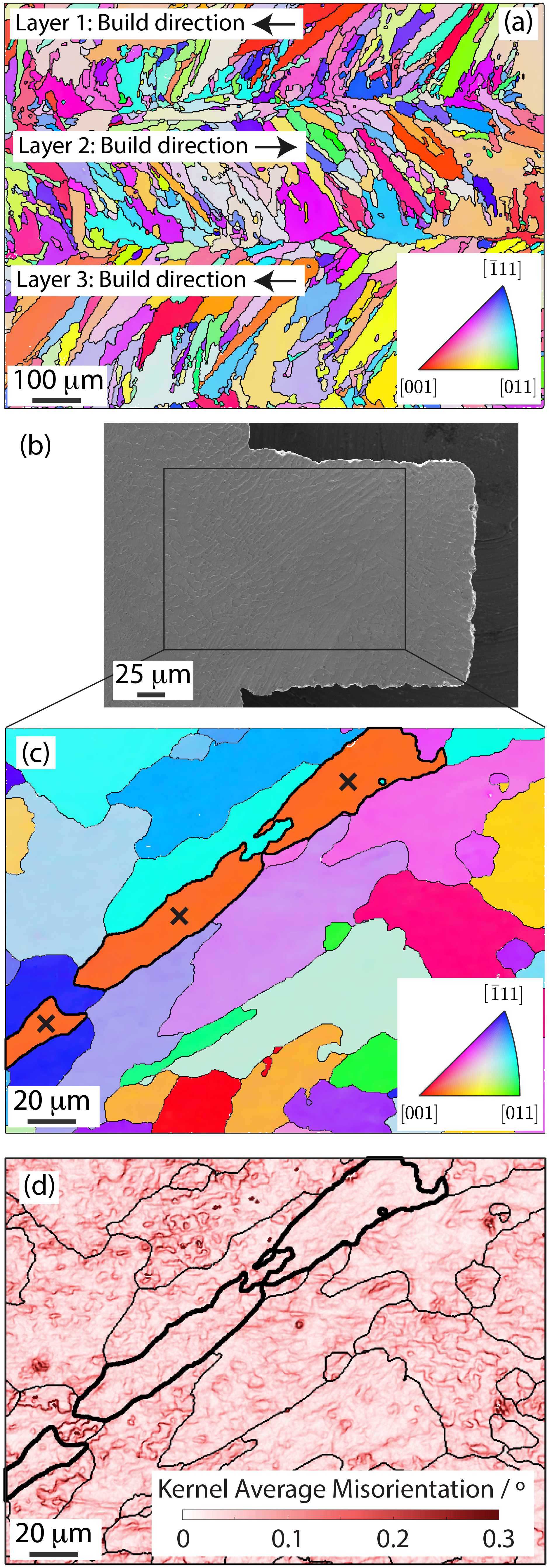}}
    \caption{From a bi-directional thin-wall sample manufactured via DED-LAM: (a) an IPF-Z orientation map obtained using EBSD. An SE image of the sample examined by DFXM is shown in (b), from which (c) an EBSD, IPF-Z map was obtained at high angular and spatial resolution, with  `$\times$' denoting the surface breaking grain observed by DFXM. In (d) the Kernel Average Misorientation (KAM) map from the EBSD dataset is shown.}
    \label{fig:ebsd}
    \vspace{-5 mm}
\end{figure}

\begin{figure*}[h]
    \centering
     \resizebox{1.8\columnwidth}{!}
         {\includegraphics{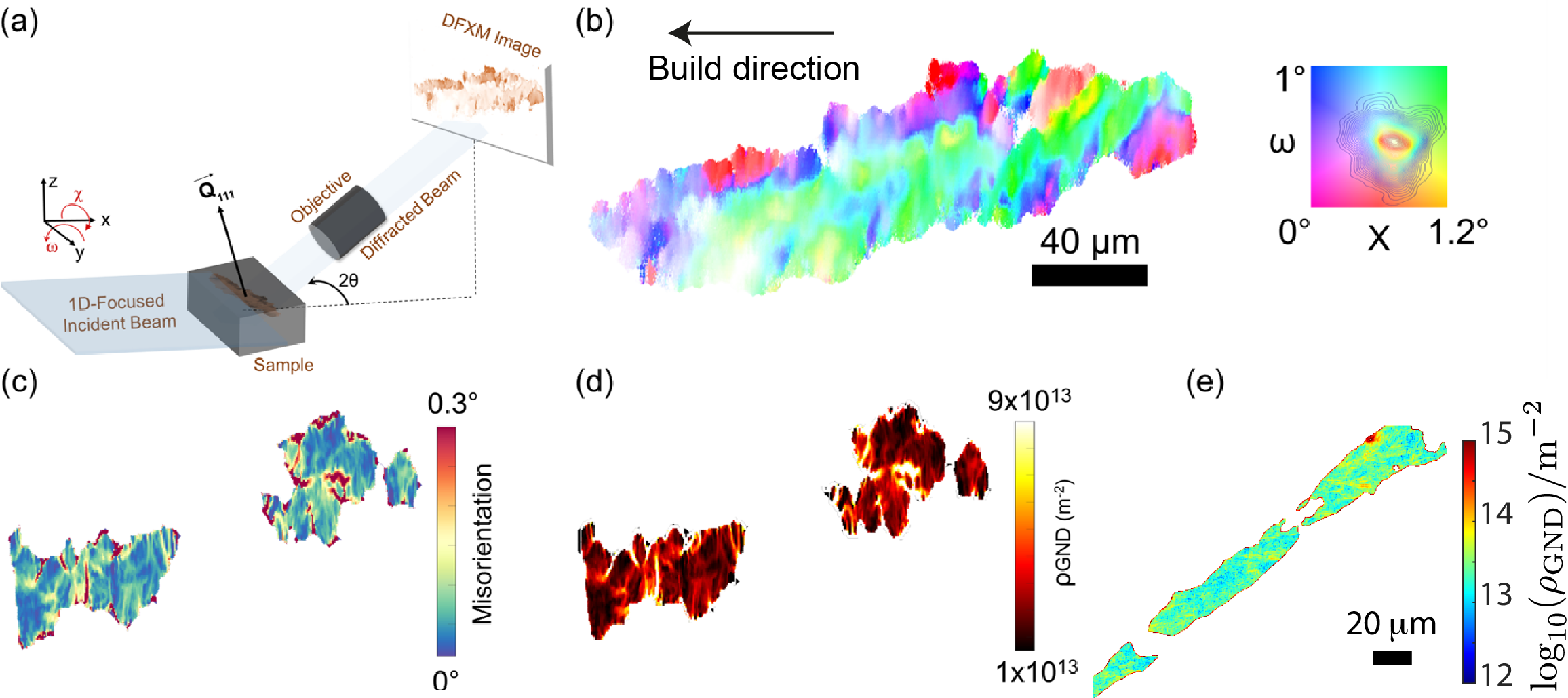}}
           \vspace{-1 mm}
    \caption{Synchrotron Dark Field X-ray Microscopy (DFXM) characterization of a Laser Additive manufactured grain. (a) Schematics of DFXM. (b) DFXM mosaicity map of a layer 30 \un{\mu m} below the surface of the grain,  with the inset showing the local 111 pole ($2\theta=20.36 \degree$) figure color key of the two sample tilts ($\omega$ and $\chi$). The mosaicity scans at the constant $2\theta$ angle reveals the spatial variation of the orientation of the lattice around the 111 scattering vector. (c) DFXM misorientation map of the surface layer emphasizing the boundaries and cells (d) calculated GND density map. The length scale in (b) applies to (c) and (d). (e) GND density map from the EBSD measurements. }
    \label{fig:dfxm}
    \vspace{-5 mm}
\end{figure*}

A bi-directional thin-wall sample of ABD-900AM nickel superalloy \cite{tang2020effect}, manufactured using DED-LAM, was sectioned in the middle perpendicular to the build plane for EBSD analysis. The sample was prepared via an abrasive metallographic route, finishing with colloidal silica. Figure~\ref{fig:ebsd}(a) shows the EBSD micrograph revealing an overview of the additively-manufactured columnar grains that follow the heat transfer direction. Distinctive grain morphology differences can be observed throughout each build layer. A cantilever (Figure~\ref{fig:ebsd}(b)) was prepared from the top layer of the sample with the following procedures: Electro Discharge Machining (EDM) was firstly used to section the specimen into a slice of $11 \times 6 \times 0.3$ mm$^3$(w/h/t) along the orientation parallel to the build direction. The slice was then mechanically ground to 200\,$\upmu$m in thickness.
Laser micromachining was then employed to section a foil of $2 \times 2.5$\,mm$^2$ to form a square cantilever of $200 \times 200$ mm$^2$. Laser damage, if any, was subsequently removed by further electropolishing using 85\,wt\% phosphoric acid (H$_3$PO$_4$) in an aqueous solution with a DC voltage of 3\,V for 60\,s. Prior to the EBSD data collection, the material was ion polished using a Gatan PIP2 at 8\,kV for 24\,mins. The incident beam and the sample surface was positioned at 8$\degree$.

The sample was characterized by EBSD using a Zeiss Merlin field emission gun scanning electron microscope (FEG-SEM). Two magnifications were acquired for the cantilever, at a step size of 0.73 \& 0.45 $\upmu$m. The higher resolution EBSD patterns (Figure~\ref{fig:ebsd}(c)) were used to infer the geometrically necessary dislocation (GND) density and residual elastic strain maps using an in-house developed MATLAB code. Here, the diffraction patterns were analyzed using a cross-correlation-based method. More details about the HR-EBSD method and its mathematical basis be found elsewhere \cite{WILKINSON2012366,Britton2013,WMG2006_1,WILKINSON2006307}.

The DFXM experiments were carried out at the beamline ID06-HXM at the European Synchrotron Facility \cite{Kutsal2019}. An incident monochromatic beam with a photon energy of 17\un{keV} was focused in the vertical direction on to the sample using a Compound Refractive Lens (CRL). The beam profile on the sample was $\approx$$200 \times 0.6 \un{\upmu m^2}$ (FWHM) in the horizontal and vertical directions, respectively. The horizontal \textit{line beam} illuminated a single plane that sliced through the depth of the crystal, defining the \textit{observation plane} for the microscope. A detector comprising a scintillator crystal, a visible microscope, and a $2160 \times 2560$ pixel$^2$ PCO.edge sCMOS camera was located 5300\,mm away from the sample. This camera had exchangeable 10$\times$ and 2$\times$ optical objectives for higher resolution and larger field of view, respectively. The diffracted beam was focused and the image was magnified by an X-ray objective lens (2D CRLs). The objective CRLs provided an image of the diffracting grain onto the far-field detector, with an X-ray magnification of $M_{\text{x}}=17.9 \times$, leading to a spatial resolution of $\approx$ 125\,nm using the 10$\times$ objective. To obtain 3D information, DFXM images were collected in 2D layers, scanning the sample in the vertical direction, $z$, to resolve the variation along the height of the crystal with 1\un{\upmu m}/step for scans using the 10$\times$ objective, and 3\un{\upmu m}/step for scans using the 2$\times$ objective, respectively.
The orientation and strain maps were collected as 2D meshes of the sample tilts and the $2\theta$ around the 111 scattering vector (Figure~\ref{fig:dfxm}(a)). The DFXM data were analyzed using the \textit{darfix} and in-house built MATLAB scripts \cite{darfix}. The matching of the GOI for EBSD and DFXM measurements was enabled by a DCT scan on the sample (see the Supp. Mat. for details).

\par 

A typical DFXM mosaicity map (a mesh of two sample tilts, $\omega$ and $\chi$) is shown in Figure~\ref{fig:dfxm}(b) from an interior slice (30\,\un{\upmu m} below the surface) of the GOI. The overall shape of the grain is elongated, which correlates with the EBSD observations. At the same time, the DFXM results display unique features in the LAM microstructure when compared to the EBSD measurements. Within the mapped layers, elongated band-like structures with similar local orientations are observed to align parallel to each other. Those structures most likely to follow the laser traverse/heat transfer direction of the manufacturing process. The overall angular spread of the GOI is less than 1.2$\degree$. The band-like structures aligned across the length of the grain show $\approx$ 0.3-0.5$\degree$ misorientation from one another, while in the interior they consist of cells with diameters <5\,$\upmu$m. Fig.~\ref{fig:dfxm} (c) shows the map of the misorientation; defined by $\Delta\gamma=\sqrt{(\Delta\omega)^{2}+(\Delta\chi)^{2} }$ \cite{yildirim20224d, yildirim2022multiscale}. Here, $\Delta \omega$ and $\Delta \chi$ are the differences between the local sample tilt center-of-mass and their grain averages \cite{darfix}. Figure~\ref{fig:dfxm}(c) and (d) show $\approx$5\,\un{\upmu m}-sized subgrains having boundaries with high dislocation densities at a layer 3\,$\upmu$m deep below the surface of the grain. The calculated dislocation densities using DFXM are found to be on the same order of magnitude with the EBSD measurements shown in Figure~\ref{fig:dfxm}(e).

\begin{figure}[t]
    \centering
    \resizebox{0.7\columnwidth}{!}
    {\includegraphics{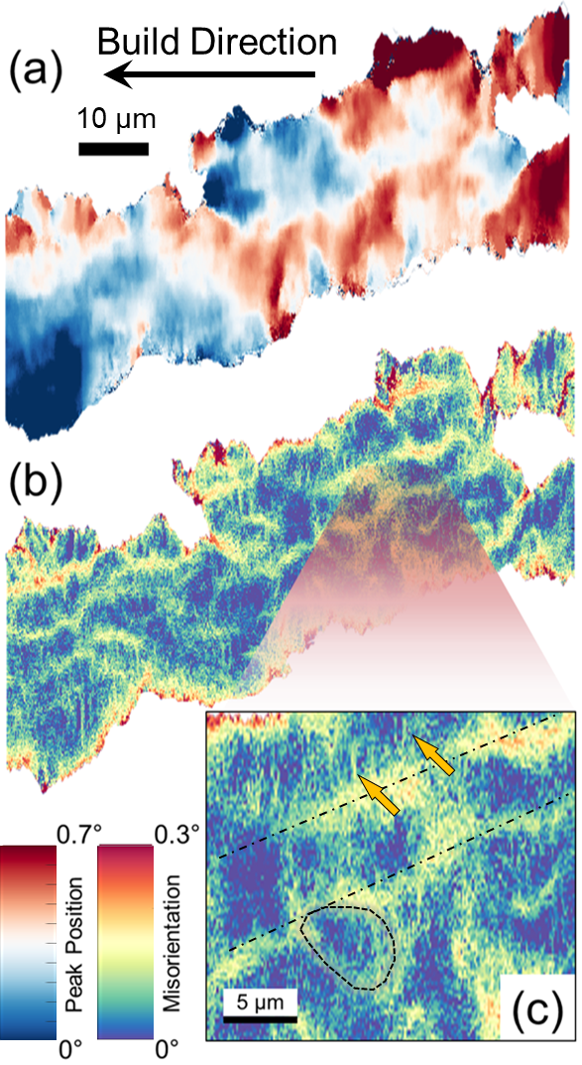}}
    \vspace{-1mm}
    \caption{(a) DFXM sample orientation $\chi$ peak position center of mass map of a given layer. The color scale range is chosen such that the contrast is maximized. (b) Calculated high resolution (using the 10$\times$ objective) misorientation map of the same layer. (c) Magnified region of the misorientation map emphasizing the cell boundaries.  }
    \label{fig:misor}
    \vspace{-5 mm}
\end{figure}

Probing deeper into the bulk, Figure~\ref{fig:misor} shows a higher resolution DFXM orientation maps of a layer embedded 30\,$\upmu$m below the surface of the sample, collected using the 10$\times$ objective. In Figure~\ref{fig:misor}(a), the $\chi$ center-of-mass map shows distinct subgrains separated with low angle grain boundaries. The overall angular spread in this map reaches up to 1$\degree$. Looking at the misorientation map (Figure~\ref{fig:misor}(b)), boundaries along the length of the grain (the band-like structure) with $\approx$0.15$\degree$ misorientation are observed. The magnified image in Figure~\ref{fig:misor}(c) shows two of these structures (denoted by dashed lines) around 5\,$\upmu$m apart (one cell is marked with the dashed shape). For comparison, the Kernel Average Misorientation (KAM) from the EBSD dataset had a mean and standard deviation of $0.045^\circ$ and $0.028^\circ$, respectively (see Figure~\ref{fig:ebsd}(d)).

\begin{figure*}
\centering
    \resizebox{1.9\columnwidth}{!}
    {\includegraphics{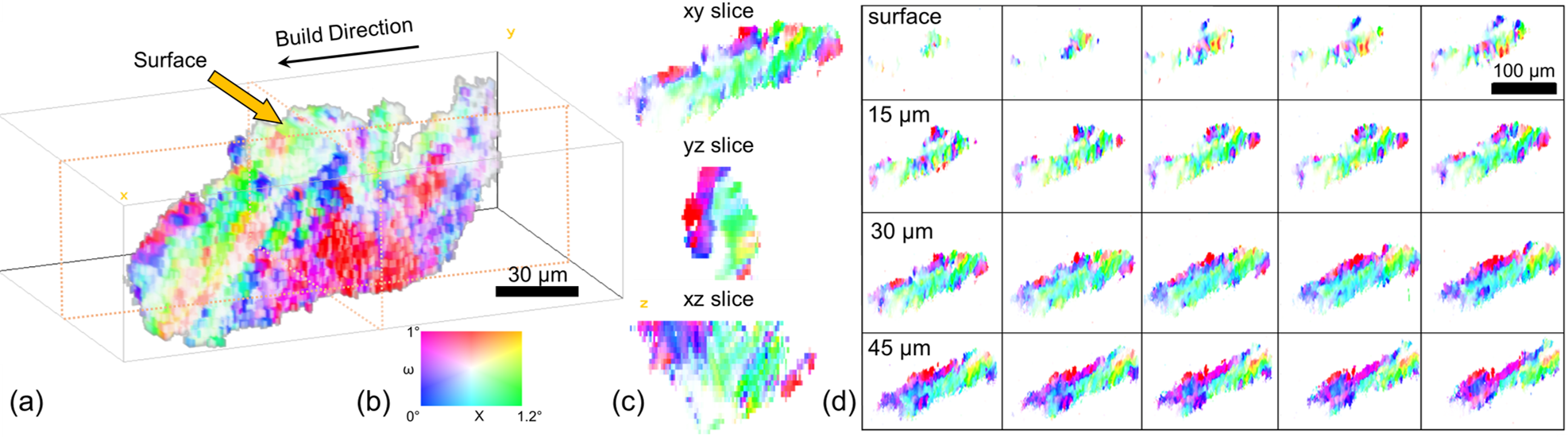}}
    \caption{3D DFXM mosaicity maps of the LAM GOI.  (a) 3D reconstruction of 2D layers showing the morphology of the grain. (b) Orientation colour key around a local 111 pole figure for the mosaicity maps. (c) different cut slices from the the 3D structure at the mid layer shown with dashed lines in the 3D reconstruction. (d) (a) 2D DFXM mosaicity maps show the structural heterogeneity across the depth of the grain from surface to interior. Each slice is 3\,$\upmu$m deeper than the previous slice. }
    \label{fig:mosaicity}
    \vspace{ -2 mm}
\end{figure*}
It is suspected that these intragranular structures may be a result of a combination of the stresses that build up due to the high cooling rates and the chemical segregation (solutes and carbides) during the fabrication \cite{raabe}. In an earlier study on the same alloy (ABD-900AM), it was shown that Ti atoms segregate to local cell boundaries at a smaller scale ($\approx$750\,nm-sized cells) in the as-fabricated state \cite{tang2023multi}. When looking closely at our DFXM results, similar boundaries can be seen as lines of 0.1$\degree$ misorientation $\sim$400-800\,nm range between (see yellow arrows in Figure~\ref{fig:misor}(c)). 

Going beyond the complex microstructures in a given 2D layer within the volume, we now turn to the 3D structure of the GOI in Figure~\ref{fig:mosaicity}. The 3D structure of the grain displays significant heterogeneity in the local orientation. We observe orientation bands extending from the surface towards the interior of the grain, spanning more than 50 $\mu$m. All three of the 2D slices, ($xy$, $yz$ and $xz$, Figure~\ref{fig:mosaicity}(c)), from the 3D volume shows rather sharp orientation changes, compared to smooth and gradual changes observed in highly-deformed metals \cite{yildirim20224d}. This may be due to the pinning effect of the chemical segregation on the dislocations, making them less mobile during the cooling thus creating well-separated boundaries with measurable orientation differences less than 0.5$\degree$. Recently, using DFXM we found that even low concentrations (below 0.1\%) of solute elements can have a significant dragging effect on the mobility of dislocations, especially for loops, even at high temperature annealing\cite{mavrikakis2019}. For superalloys such segregation has been associated, for example, with pipe diffusion during creep \cite{KONTIS201876}. Note that all the slices shown in Figure~\ref{fig:mosaicity}(c)) have different morphologies. This indicates the cooling gradients during the fabrication are anisotropic, causing a heterogenous 3D orientation distribution and thus resulting in anistropic mechanical properties \cite{munoz2016effect, tang2023multi}.


Similar to the single-slice observation in Figure~\ref{fig:dfxm}(b), distinct band-like cell structures with similar orientations exist in all slices in Figure~\ref{fig:mosaicity}(d). Inside the bands, an orientation distribution is also observed in the shape of cell structures. These structures are visible thanks to the high angular resolution of the DFXM, which can reveal subtle orientation variations that are otherwise impossible to observe with EBSD \cite{hlushko2020dark}. These variations near the sub-grain boundaries are likely due to the presence of a high dislocation density, specifically geometrically necessary dislocations (GNDs) caused by large lattice distortions. The local orientation distributions over the measured volume are far from homogeneous (see  Supplementary Figure S3.) This is attributed to the complex cooling gradients during the manufacturing.  

\begin{figure}[!h]
    \centering
     \resizebox{0.98\columnwidth}{!}
     {\includegraphics{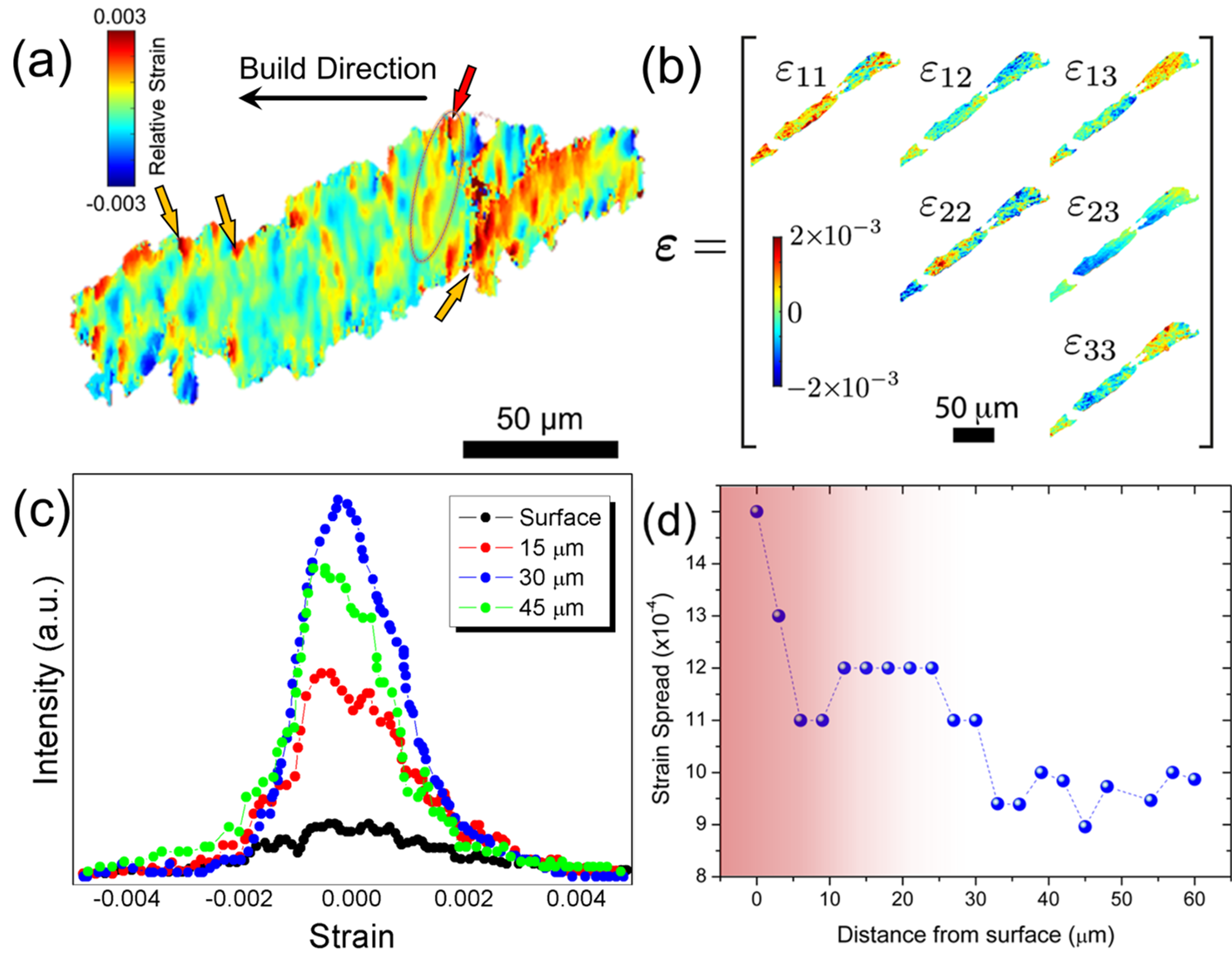}}
    \caption{DFXM relative strain maps of a LAM grain. (a) DFXM relative axial strain map of a slice ~30\,\un{\mu m} below the surface of the grain. (b) Calculated strain tensor components from the EBSD measurements. (c) DFXM-measured strain distribution at different depths of the grain). (d) Total DFXM-measured strain spread as a function of distance from the surface of the grain. }
    \label{fig:strain}
\end{figure}

The observed lattice distortions should also be considered in connection with the residual stresses. The spread of relative elastic strain of a given layer is shown in Figure~\ref{fig:strain}(a), demonstrating zones that have highly-accumulated residual strain. The measured strain component is perpendicular to the [111] direction with $\approx\,10.15\degree$ offset. At a first glance, the strain distribution is fairly distinctive around the cell structure and the sub-grains, which matches the observation from the mosaicity and misorientation distribution. The misorientation maps suggest that a significantly high dislocation content creates the cell structures, with distinct orientation and strain states. A closer look at the DFXM strain map in Figure~\ref{fig:strain}(a) reveals zones with alternating compressive and tensile strain. These $\approx$5\,$\upmu$m-sized zones correspond well to the cell structures with distinct local orientations, indicating a coupling of the lattice distortion and the d-spacing change in the interior of the grain. However, the finer structures which may be linked to the solute segregation (i.e. Ti \cite{tang2023multi} in Figure~\ref{fig:misor}(c)) are not visible in the strain map. Instead, a rather homogeneous strain distribution is observed within these cells that are aligned along the length of the grain with alternating strain states. Moreover, high strains (>0.003) are observed around the grain boundaries (marked with yellow arrows). This is attributed to the constraining effect of the neighbouring grains during the solidification, creating stress concentration zones around the grain boundaries. Using the \{111\} diffraction elastic constant of a similar alloy\cite{aba2016determination}, it is found that these zones may have stresses extending to 850\,MPa; on the order of the yield stress of the alloy\cite{tang2020effect}(See the Supplementary Figure S4). In addition, the strain fields of these highly-stressed zones propagate towards the interior of the grain, increasing the overall residual stress level (marked by the red arrow and the ellipse). The stored energy created by the high residual stresses and the high dislocation density can trigger static recrystallization and should be considered for subsequent anisotropy-removing recrystallization and $\gamma'$ transformation heat treatments. 

Figure~\ref{fig:strain}(b) shows the residual elastic strain tensor obtained from the EBSD measurements, in the sample reference frame. All components show significant deformation patterning within the GOI. These form cell type structures of compressive or tensile strain. This cell size approximately matches those observed in the DFXM strain map in Figure~\ref{fig:strain}(a), indicating these features can be resolved by both techniques. Although the measured strain magnitudes are similar between the two techniques, EBSD maps do not reveal the clear alternating tension and compression patterning. This may be due to the strain relaxation at the surface. Figure~\ref{fig:strain}(c) shows a heterogeneous distribution of the lattice strain within the measured volume. Although the magnitude of the strain decreases at the surface where relaxation is expected, the strain spread shown in Figure~\ref{fig:strain}(d) shows an increase closer to the surface (region marked by the red color). This deviation from the surface to the interior ranging over some tens of micrometers is likely to be due to the polishing effects on the surface when the sample was prepared for the EBSD measurements prior to the DFXM experiment.

A clear picture now emerges of how the observed complex intragranular microstructures are formed via DED-LAM technique in the ABD-900AM alloy. During solidification, large volume changes that occur due to the phase transformation giving rise to the generation of many dislocations in order to accommodate the volume differences \cite{debroy2018additive}. These dislocations are predominantly GNDs, creating intragranular cells with low angle grain boundaries, manifesting themselves as measurable misorientation in the mosaicity scans. Due to the limited time spent at high temperatures during LAM, small cells with various sizes below 5\,$\upmu$m are formed, unlike their cast counterparts\cite{jahangiri2012study}. The cell sizes remain small because of the low mobility of the dislocations caused by the combined effect of the rapid cooling and the chemical segregation of the solute atoms and carbides\cite{tang2023multi}. The DFXM strain scans show that the effective d-spacing changes during solidification, resulting in prominent stresses around the grain boundaries as a result of the contact with neighboring grains during the fabrication process. Some of these stresses extend  tens of micrometers towards the interior of the grain. As the strain distribution within the intragranular cells are homogeneous with clear alternation of tension and compression zones from cell to cell, we argue that the strain distribution within the cells are dominated by the thermal effects rather than the chemical effect. 

The present study demonstrated sub-surface variations of the local strain and orientation of a surface breaking grain in an additively-manufactured sample using DFXM. The DFXM method successfully resolves complex 3D networks of a highly-strained grain in the as fabricated state, non-destructively revealing the deformation substructure patterning that cannot be resolved by other techniques in bulk. We compare the DFXM results to the EBSD measurements of \textit{the same} grain. Our results show that band-like structure aligned along the length of the grain and $\approx$5 $\mu$m-sized cell structures within these bands are formed due to the extreme thermal gradients during LAM, and that they are aligned towards the build direction along the length of the grain. At a finer length scale, misorientation lines of 400-800\,nm connecting larger cell structures are observed. These lines were attributed to the chemical segregation, based on a recent study on the same alloy\cite{tang2023multi}. 3D DFXM strain and orientation maps show that distortion patterning and magnitudes within an individual grain vary significantly (on the order of several micrometers). This demonstrates that surface-based characterisation techniques, such as EBSD, provide measurements that are not representative of complex sub-surface microplasticity for materials fabricated via LAM. Our DFXM results provide unprecedented 3D intragranular information on the lattice strain and distortion, opening up new avenues not only for potential improvements in the design of heat treatment routes and future experiments for \textit{in-situ} manufacturing, but also for new input parameters for modelling such as geometrical boundary conditions, misorientation angles and strain fields.


    



\section*{Acknowledgements}
We thank the ESRF for provision of beamtime at ID06-HXM and ID11. YC acknowledge the support from the RMIT Vice Chancellor's Senior Research Fellowship. PJW and YT acknowledge support from the European Research Council Grant No 695638. The authors are grateful for Alloyed Ltd for provision of material. This research used resources of the Advanced Photon Source, a U.S. Department of Energy (DOE) Office of Science user facility at Argonne National Laboratory and is based on research supported by the U.S. DOE Office of Science-Basic Energy Sciences, under Contract No. DE-AC02-06CH11357.

\bibliography{ref_scripta}

\end{sloppypar}
\end{document}